\documentclass[aps,pre,twocolumn,superscriptaddress,nofootinbib,tightenlines,10pt]{revtex4-2}

\usepackage{graphicx}
\usepackage{bm}
\usepackage{amsmath}
\usepackage{amssymb}
\usepackage{anyfontsize}
\usepackage{colortbl}
\usepackage{xcolor}
\usepackage{nicefrac}

\definecolor{myblue}{HTML}{0466c8} 
\definecolor{mygreen}{HTML}{00693e}
\definecolor{myred}{HTML}{9d162e}
\usepackage[
colorlinks = true,
linkcolor  = mygreen,
citecolor  = myblue,
urlcolor   = myblue,
bookmarksnumbered = true,
]{hyperref}

\begin{document}

\title{Disentangling intermittent flow structure contributions to anomalous scaling and multifractality in turbulence}

\author{Ritwik Mukherjee}
\email{ritwikmukherjee.kh@gmail.com}
\affiliation{International Centre for Theoretical Sciences, Tata Institute of Fundamental Research, Bengaluru, Karnataka 560086, India}

\author{Siddhartha Mukherjee}
\email{smukherjee@iitk.ac.in}
\affiliation{Department of Mechanical Engineering, Indian Institute of Technology, Kanpur, Uttar Pradesh 208016, India}

\date{\today}

\begin{abstract}
Intermittency in turbulence manifests as intense vortices and sharp peaks of dissipation. Causing the breakdown of Kolmogorov's simple self-similar theory, it leads to anomalous scaling, multifractality and so far remains beyond the scope of a complete theoretical description. How intermittent flow structures influence these different measurements is not known quantitatively. With a simple filtering procedure---thresholding vorticity and inverting the Biot-Savart law to generate filtered velocity fields---we show the effects of intermittent flow structures can be disentangled. As extreme vorticity contributions to the velocity field are filtered out, the energy spectrum scaling persists, while the bottleneck is flattened, and structure function scalings tend towards their Kolmogorov values. The approach is more rapid for transverse exponents, revealing the selective importance of intensely swirling flow regions. Similarly, the extent of multifractality reduces as intermittency is filtered, shrinking the range of roughness singularity exponents. The residual fields are curiously more multifractal, but their structure begins to break away from an underlying turbulence skeleton. The effects on vortex stretching and strain self-amplification are quantified. Our work shows that a Biot--Savart approach can selectively remove the effects of intermittency from turbulence, and hence from its scalings.
\end{abstract}

\maketitle

\section{Introduction}\label{Intro}
Perhaps the most distinctive and readily observable feature of turbulence is its extreme patchiness and bursts in the magnitude of certain observables like vorticity and dissipation~\cite{dhawan1958some,Moffatt,Frisch_1995,kleckner2013creation, saw2016experimental,yeung2015extreme,buaria2026turbulence}. This can be traced back to some of the dynamics at the heart of turbulence, like strain self-amplification, vortex-stretching and reconnections, and the multiplicative accumulation of energy dissipation over the cascading process~\cite{she1990intermittent,Moffatt_Kida_Ohkitani_1994,kida1994vortex, tsinober2009informal,Carbone_Bragg_2020,Johnson_2021,debue2021three, yao2022vortex,matsuzawa2023creation}. This phenomenon of patchiness and its effects on the statistics of turbulence together are termed \textit{intermittency}. Tell-tale observations of intermittency are many. In already second order statistics like energy spectra, intermittency can be felt upto the inertial range as corrections to the spectral scaling, while the bottleneck effect reflects small scale accumulation of energy which may be related to intense flow structures~\cite{kaneda2003energy,frisch2008hyperviscosity,donzis2010bottleneck}. Intermittency affects higher order statistical moments more strongly, leading to the anomalous scaling of velocity structure functions~\cite{anselmet1984high,Sreeni}. This is the deviation of $ \langle |\delta u(r)|^p \rangle \sim r^{\zeta_p}$ from the self-similar Kolmogorov scaling moments $\zeta_p = p/3$ to a non-linear convex form. Moreover, the nature of scaling in longitudinal and transverse velocity projections are different. Anomalous scaling ties velocity structure functions to the spatial variation of energy dissipation in terms of local H\"older exponents $h$. While the intermittency-free Kolmogorov theory predicts a single scaling exponent for the entire flow, $h=1/3$, leading to the $\zeta_p = p/3$ prediction assuming self-similarity, real turbulence appears to allow a spread of scaling exponents centered around the Kolmogorov prediction. This is now understood from multifractality of the dissipation where there can be a range of exponents $h_{\rm min} < h < h_{\rm max}$~\cite{Frisch-Parisi,MS-Nucl,frisch1991global,sreenivasan1991fractals,boffetta2008twenty,mukherjee2024turbulent}, as opposed to the monofractal classical view $h=1/3$. While several aspects of intermittency are now well understood separately, there is still no overarching theoretical framework that explains all these effects from first principles, making this a standing challenge in turbulence theory.

It is now believed that \textit{intense vorticity} is key to the observed effects of intermittency~\cite{she1990intermittent,she1994universal,buaria2019extreme,debue2021three,buaria2026turbulence}. Unbounded vorticity growth is also the metric used as a proxy to search for finite-time blowup of Navier--Stokes solutions~\cite{beale1984remarks}. It is intriguing to consider what this essentially means. On the one hand, there is the classical Kolmogorovean picture of turbulence, which by comparison to actual turbulence, appears to describe a phenomenon that is quiescent and well-behaved while essentially being multiscale, irregular and even rough. On the other hand, bursts of intermittency render all these (relatively) simpler and well-behaved aspects of turbulence moot by bending scaling laws, stretching distributions and adding a complex singularity structure. Real turbulence seems to be a superposition of both of these states.

It is compelling to ask how exactly intermittent flow structures affect statistics and scalings in turbulence. To address this, a procedure to filter out intermittency needs to be adopted, and there is no obvious route or a singular approach to this question. There have been very few studies that have dealt with this issue. Notably, there is the surgical procedure called the decimated Navier--Stokes, which imposes dynamical constraints on Fourier-modes while evolving the solution on a fractal Fourier set~\cite{Frischetal2012,buzzicotti2016lagrangian,Ray2018}. The consequent inhibition of certain wavector triadic interactions leads to a flow that becomes increasingly non-intermittent as the fractal dimension of decimation is increased. The solutions of the projected dynamical system lose both inertial-range and small-scale intermittency and recover Kolmogorov scaling~\cite{kankaria2026reduction}. This process generates non-intermittent turbulence by altering the global dynamics itself, and the procedure does not allow for singling out intermittency at a chosen scale. Wavelet based approaches to filtering intermittency in atmospheric boundary layers~\cite{katul1994intermittency} have also been proposed, which allow identification and artificial suppression of intermittent structures and their effects. Recent work on identification and disentanglement of coherent structures in turbulence adopted an approach using the Helmholtz-decomposition~\cite{mukherjee2022correlation}, which allows reconstruction of integral fields (like velocity) from its gradient fields (like dilation and vorticity). This allows a more direct, kinematic approach to separate out vorticity contributions in inducing the velocity field in a Biot--Savart sense. Such an approach was also adopted to unravel the alignment of vorticity with the intermediate strain eigendirection~\cite{hamlington2008local}.

In this work, we adopt a Biot--Savart approach to filtering intermittent, extreme vorticity, hence removing strongly swirling velocity regions around vortex cores from the velocity field, while retaining the background induced velocity structures. This process allows disentangling the velocity field from its superposition in a Biot--Savart sense. We then study the scaling in filtered velocity fields over different thresholds, quantifying the deviation and approach towards Kolmogorov scaling, by computing structure functions and multifractality. We also study the residual velocity field, which has been filtered out, to study a field that is purely the intermittent part of turbulence. Our approach allows us to cleanly separate out the effects of intermittency on the velocity field, and we show that the true background field in turbulence is Kolmogorovean. We quantify the approach to simple scaling behaviour over our filtering magnitude, and discuss the role of vortices, vortex-stretching and strain self-amplification in making turbulence increasingly intermittent.

\begin{figure*}
    \centering
    \includegraphics[width=0.99\linewidth]{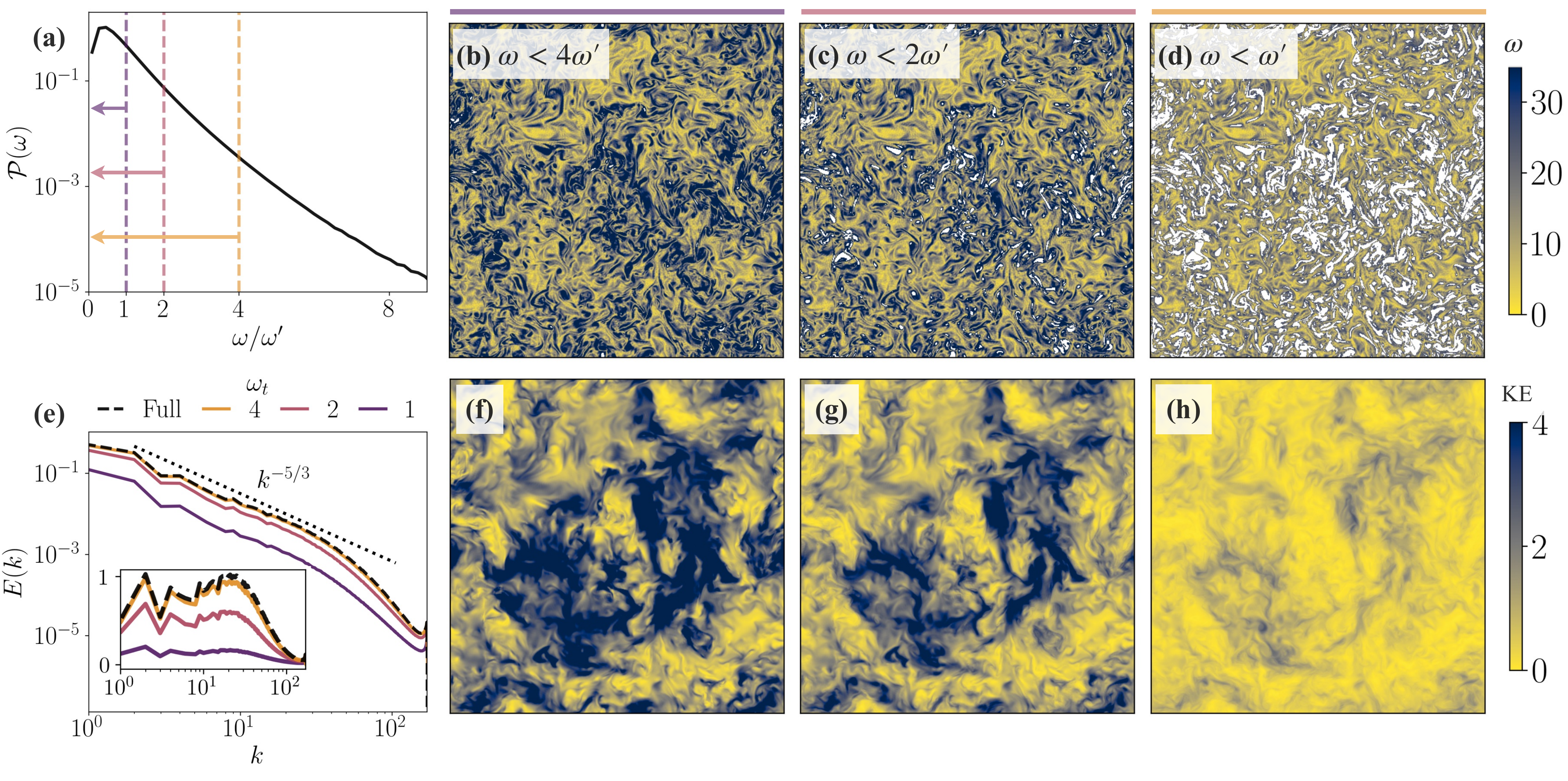}
    \caption{\textbf{Filtering intermittency.} 
    \textbf{(a)} Probability distribution of the vorticity magnitude $\omega$. Three levels of thresholding $\omega<\omega_t$ have been marked, namely with $\omega_t = \lbrace 4\omega^\prime,2\omega^\prime,\omega^\prime\rbrace$. Panels \textbf{(b)}-\textbf{(d)} show increasingly thresholded vorticity fields, where values above the threshold and have been masked out (most prominent in \textbf{(d)}). Cross-sections of the kinetic energy of reconstructed velocity fields ${\rm KE} = \nicefrac{1}{2}|\widetilde{\bf u}|^2$ are shown in \textbf{(f)} to \textbf{(h)}. Curiously, the velocity field retains much of its structure even upto a thresholding $\omega_t = \omega^\prime$. This is also reflected in \textbf{(e)} the energy spectrum of the reconstructed velocity fields. In comparison to the spectrum of the full velocity field, the reconstructed velocity spectra show a drop in magnitude, while retaining the same $k^{-5/3}$ scaling over an almost identical range of wavenumbers. Inset shows compensated spectra $k^{5/3}E(k)$ on a log-linear scale reflecting a flattening of the bottleneck.}
    \label{fig1}
\end{figure*}

\section{Methods}\label{Methods}
We consider the three dimensional incompressible Navier--Stokes equations for a velocity field $\mathbf{u}(\mathbf{x},t)$
\begin{equation}
    \partial_t \mathbf{u} + (\mathbf{u} \cdot \nabla)\mathbf{u} = -\nabla p + \nu \nabla^2 \mathbf{u} + {\mathbf{f}}, \qquad \nabla \cdot \mathbf{u}= 0,
\end{equation}
where $p$ is the pressure and $\nu$ is the kinematic viscosity and $\mathbf{f} $ is large scale body forcing. Vorticity $\boldsymbol{\omega}$ is defined as $\boldsymbol{\omega} = \mathbf{\nabla} \times \mathbf{u}$ and $\omega=|\boldsymbol{\omega}|$. Due to the lack of far-field/boundary fields over periodic domains, and further due to incompressibility, the Helmholtz-decomposition reduces to the simpler Biot--Savart law which is simply a kinematic relation between the velocity and vorticity fields, given as
\begin{equation}
{\bf u}({\bf x},t)
=
\frac{1}{4\pi}
\int_V
\boldsymbol{\omega}({\bf x}',t)
\times
\frac{{\bf x}-{\bf x}'}
{|{\bf x}-{\bf x}'|^3}
\,{\rm d}{\bf x}'.
\label{eq:biot-savart}
\end{equation}
The velocity field at any given point has non-local contributions from the vorticity \textit{everywhere} in space. This additive nature of the Biot--Savart law, therefore, allows reconstructing velocity fields corresponding to only a chosen range or spatial region of the vorticity field, a route to disentangling the effects of intermittent vorticity on the velocity field. 

We create thresholded vorticity fields
\begin{equation}
    \widetilde{\boldsymbol{\omega}}({\bf x}) = \boldsymbol{\omega}({\bf x})|_{\omega < \omega_t},
\end{equation} 
where $\omega_t$ is a threshold taken to be some level of the rms vorticity of the full field, and regions with $\boldsymbol{\omega}({\bf x})|_{\omega \geqslant \omega_t}$ are masked to zero. The reconstructed velocity field $\widetilde{\bf {u}}({\bf x})$ is obtained by inverting the relation $\Delta \widetilde{\bf {u}}({\bf x}) = -\nabla \times \widetilde{\boldsymbol{\omega}}({\bf x})$ spectrally. Similarly the residual velocity field can be constructed by inverting the vorticity in the masked regions to obtain $\widetilde{\bf u}^{R}$.

The incompressible Navier--Stokes equations are solved using a pseudo-spectral method~\cite{gottlieb1977numerical} subject to constant power injection~\cite{pope2004forcing}, over a $512^3$ simulation domain with 2/3-dealiasing. We compute the Taylor microscale $\lambda = u_{\mathrm{rms}} \sqrt{15 \nu /{\overline{\varepsilon}}}$, where, $u_{\mathrm{rms}}$ is the root-mean-square velocity, and $\overline{\varepsilon}$ is the mean energy dissipation rate. The corresponding Taylor microscale Reynolds number is defined as $Re_{\lambda} = u_{\mathrm{rms}} \lambda / \nu$. Simulations were performed across a range of Taylor microscale Reynolds numbers ($Re_{\lambda}$), the results presented here are based on $Re_{\lambda} \approx 200$. 

\section{Results}\label{results}
We start by showing the qualitative effects of intermittency filtering. Fig.~\ref{fig1}(a) shows the probability distribution of the vorticity magnitude $\omega$ normalized by the root-mean-square vorticity $\omega^\prime$. Our filtering approach consists of retaining vorticity \textit{below} a chosen threshold $\omega_t$ and masking out the vorticity \textit{above} the threshold. Three values of the threshold have been marked by the vertical lines in Fig.~\ref{fig1}(a), the arrows show the range of vorticity retained. Panels (b)-(d) show an increased level of filtering with $\omega_t = \lbrace 4\omega^\prime, 2\omega^\prime, \omega^\prime \rbrace$, from left to right. The white empty regions correspond to vorticity above the threshold which has been masked out. 

The filtered velocity field $\widetilde{\bf u}$ is obtained by inverting the Biot-Savart law ( see \S~\ref{Methods}) on the filtered vorticity fields. We show the kinetic energy of the reconstructed velocity fields (which we shall later refer to as simply the filtered velocity), defined as $\mathrm{KE}=\nicefrac{1}{2}|\widetilde{\mathbf{u}}|^2$ in Figs.~\ref{fig1}(f)-~\ref{fig1}(h), corresponding to each level of vorticity thresholding, respectively in Figs.~\ref{fig1}(b)-~\ref{fig1}(d). Interestingly, the reconstructed kinetic energy retains its larger scale features and the filtering operation seems to mainly reduce the amplitude of the most extreme kinetic energy, particularly when the filtering approaches the level of intermediate vorticity, $\omega_t \approx \omega^\prime$. 

Fig.~\ref{fig1}(e) shows the spectra of the filtered fields in comparison to the full velocity field. This reveals an interesting effect, that the filtering operation only reduces the magnitude of the spectra, that are seen to shift downward with increasing filtering, consistent with the field plots. The shape of the spectra seems to remain unchanged. In fact, it appears that the scaling range might have \textit{increased} by a small fraction due to the filtering. Turbulence spectra are known to be effected by bottlenecks, along with inertial range intermittency effects~\cite{kaneda2003energy,donzis2010bottleneck}, which leads to a bump in the spectra due to an energy pile-up. Our procedure not only filters out intense vorticity effects on the velocity field, but curiously flattens the bottleneck. We show compensated spectra $k^{5/3}E(k)$ on a linear scale in the inset of Fig.~\ref{fig1}(e) which confirms this effect. The velocity structures associated with strong vorticity, therefore, play a role in the perceived bottleneck effect. This interestingly opens up questions about the Kolmogorov constant, but we do not pursue these directions further in our present study.

While this effect cannot be trivially justified, it perhaps shows an interesting feature of structure organization in turbulence. It seems that vorticity upto an intermediate level, $\omega \approx \omega^\prime$, already induces a velocity field consistent with a $E(k) \sim k^{-5/3}$ energy hierarchy, showing that the background velocity field is classically turbulent. Higher levels of vorticity, $\omega > \omega^\prime$, are distributed in a manner that including their Biot--Savart contributions does not violate this scaling at small wavenumbers, until $\omega \gg \omega^\prime$ where the intensification of small vortices leads to an energy pileup at small scales, as reflected in the bottleneck bump. 

In fact, even in the case of strong vorticity filtering ($\omega_t = \omega^\prime$, which removes almost 25\% of vorticity), the filtered velocity field corresponding to $\omega < \omega^\prime$ already has the essential turbulence organization associated with a $k^{-5/3}$ Kolmogorov scaling. This reinforces that, even though convoluted, the vorticity field maintains a level of organization upto relatively mild levels of vorticity magnitude, which often is considered structureless, but clearly has a strong effect on flow organization and the energy hierarchy~\cite{mukherjee2022correlation}. 

We note that the accuracy of the velocity reconstruction is up to machine precision. We have tested this by computing the difference between the full velocity field ${\bf u}$ and the sum of the Biot-Savart reconstructed filtered and residual velocity fields ($\widetilde{\bf u} + \widetilde{\bf u}^R$), which for $\omega_t = \omega^\prime$ yields values of $\mathcal{O}(10^{-14})$.

\begin{figure*}
    \centering
    \includegraphics[width=1\linewidth]{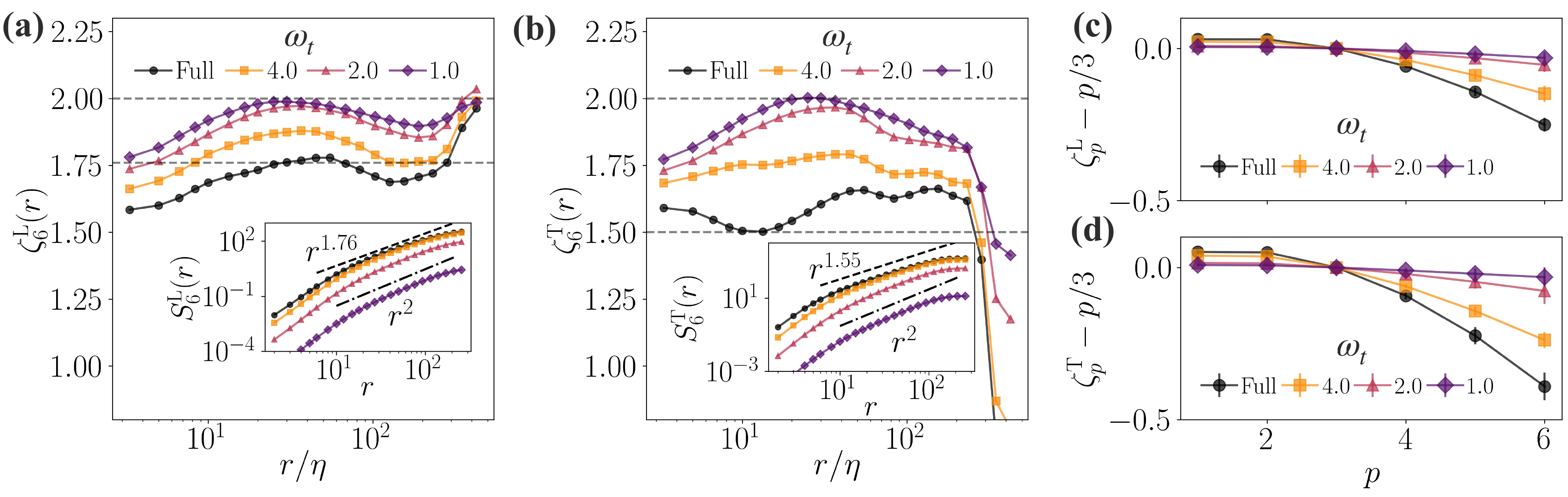}
    \caption{\textbf{Conditioned structure functions.} \textbf{(Inset a)} Longitudinal and \textbf{(Inset b)} transverse sixth-order velocity structure functions are shown for different thresholds $\omega_t$ (which is mentioned in terms of multiples of $\omega^\prime$). The main panels show the scaling exponent \textbf{(a)} $\zeta^{\rm L}$ and \textbf{(b)} $\zeta^{\rm T}$ as the local slopes of the structure functions in the inset. Both scaling exponents shift from their anomalous scaling values to the Kolmogorov prediction, with increasing intermittency filtering. The transverse exponents show a larger initial shift reflecting the role of strong vorticity is stronger for transverse anomalous scaling. The deviation of all scaling moments from the Kolmogorov prediction, for increasing intermittency filtering, is shown for \textbf{(c)} longitudinal exponents as $\zeta_p^{\rm L} - p/3$ and \textbf{(d)} transverse exponents as $\zeta_p^{\rm T} - p/3$. As intermittency filtering is increased, the deviation from Kolmogorov scaling collapses to zero.}
    \label{fig2}
\end{figure*}

What then is the effect of this filtering on intermittency? While regions of strong vorticity are considered intermittent in their own right, our procedure removes these regions in a manner which subtracts out their contribution to the velocity field. In a way, our procedure extracts turbulent velocity fields from their superposition in the Biot--Savart sense, separating the background turbulence field (corresponding to $\omega<\omega_t$) on top of which intermittent vorticity adds its contributions. We now probe the scaling behaviour of these background fields for varying $\omega_t$.

We start with the longitudinal and transverse structure functions, which are defined as: 
\begin{equation}
    S^{\rm L}_p(r) = \langle \delta u_L^p(r)  \rangle  \qquad \mathrm{and} \qquad S^{\rm T}_p(r) = \langle \delta u_T^p(r) \rangle,
\end{equation}
where $\delta u_{\rm L}(r)$ is the longitudinal velocity increment defined as
\begin{equation}
    \delta u_{\rm L}(r) = [\mathbf{u(x+r)-u(x)}] \cdot \widehat{r},
\end{equation} 
and $\delta u_{\rm T}(r)$ is the transverse velocity increment, defined as 
\begin{equation}
    \delta u_{\rm T}(r)=\sqrt{|\delta {\bf u}(r)|^2-\delta u_{\rm L}^2(r)}.
\end{equation}
Time averaging over nonequilibrium statistically stationary states is denoted by $\langle \cdot \rangle$. 

The sixth-order longitudinal ($S^{\rm L}_6$) and transverse ($S^{\rm T}_6$) velocity structure functions are shown in the insets of Fig.\ref{fig2}(a) and Fig.\ref{fig2}(b), respectively. There seems to be a clear change in the scaling behaviour of the structure functions, although the range of scaling is small. The scaling of $S^{\rm L}_6$ in Fig.\ref{fig2}(a) inset goes from $r^{1.76}$ in the full velocity field, which is the known anomalous scaling exponent, gradually upto $r^2$, which is the Kolmogorov prediction, as the filtering is increased by reducing the threshold successively from $4\omega^\prime$ to $\omega^\prime$. This shows a clear trend of intermittency effects being surgically removed. The transverse structure functions in Fig.\ref{fig2}(b) inset show the same trend, with the change that the scaling goes from $r^{1.55}$ to $r^2$.

A more clear picture of this scaling change is presented in the main panels of Fig.\ref{fig2}(a) and \ref{fig2}(b), which show the scaling exponents $\zeta^{\rm L}_6$ and $\zeta^{\rm T}_6$, respectively, calculated as the local slopes of the structure functions obtained via an Extended Self-Similarity (ESS) procedure~\cite{RayESS,BenziESS}. The full velocity field shows a local slope with a plateau around the known anomalous scaling values. As the filtering is increased, the local slope plateau gradually shifts upward. At the highest filtering shown in this figure, the plateau touches the Kolmogorov prediction of $\zeta_6^L = 2$ (naturally, the extent of the plateau is small, which is seen across turbulence data even at higher Reynolds numbers, when computing scaling exponents via local slopes~\cite{Buaria_2023_Saturation}). The transverse structure function scaling $\zeta^{\rm T}_6$ also approaches the Kolmogorov prediction. Curiously, the effects of even a small amount of filtering, for instance $\omega_t = 4\omega^\prime$, has a pronounced effect on the transverse structure functions and the scaling exponent changes significantly, an effect which is only mild for the longitudinal structure functions. 

\begin{figure*}
    \centering
    \includegraphics[width=1\linewidth]{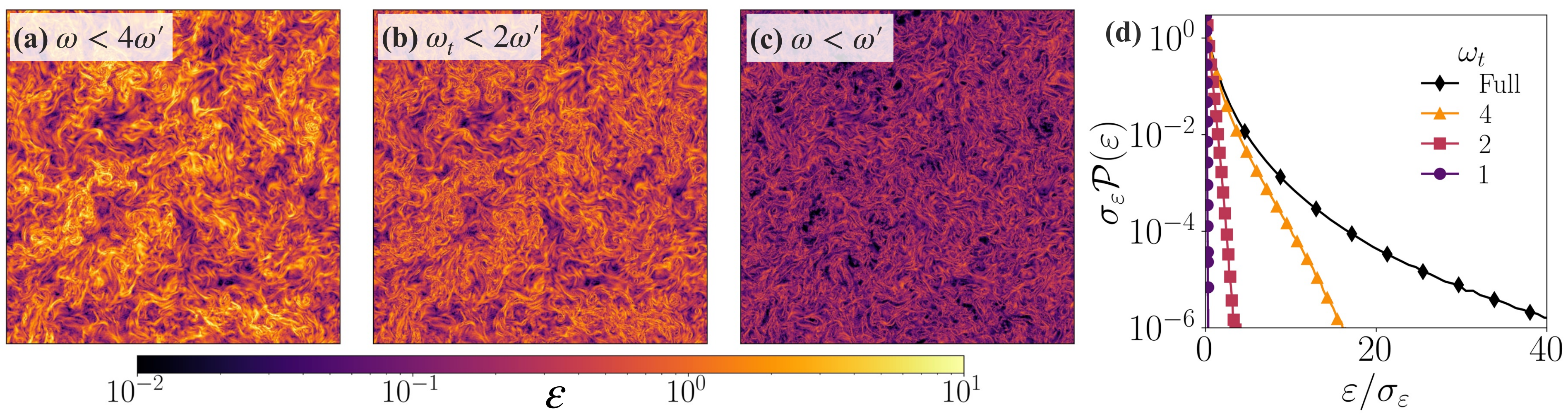}
    \caption{\textbf{Filtered dissipation}. \textbf{(a)}-\textbf{(c)} Filtered dissipation fields $\varepsilon = 2\nu\widetilde{S}_{ij}\widetilde{S}_{ij}$ corresponding to the reconstructed velocity fields $\widetilde{\bf u}$, for different levels of thresholding, have been shown at a crossection of the 3D dataset. The colours show dissipation logarithmically, with the same extent used for all the plots to keep the magnitude comparable. Interestingly, the dissipation field retains a rich structure even with the strongest filtering in \textbf{(a)} with $\omega_t = \omega^\prime$, albeit with extremely small values. The probability distribution of the filtered dissipation fields is shown in \textbf{(d)}, for different levels of thresholds $\omega_t$ (mentioned in terms of multiples of $\omega^\prime$), where the effect of filtering is seen to clearly reduce the level of dissipation to very small values.}
    \label{fig3}
\end{figure*}

This hints that regions of strong vorticity, which are surrounded by self-generated swirling flows, are crucial to the anomalous scaling of transverse structure functions that are prone to the influence of anti-aligned velocity vectors. Local regions of swirling, vortex-like motion, create such anti-aligned velocity. An analysis of this sort perhaps provides clues as to how flow structures influence longitudinal and transverse structure functions differently, also showing the importance of the different constructions of structure functions in turbulence.

Next, we quantify the deviations from Kolmogorov scaling of structure functions of orders upto 6, by considering the quantities $\zeta^{\rm L}_p - p/3$ in Fig.\ref{fig2}(c) and $\zeta^{\rm T}_p - p/3$ in Fig.\ref{fig2}(d). The scaling exponents of the full unfiltered velocity field deviate most strongly, as expected, and there is a clear approach to Kolmogorov scaling of $p/3$ as the intermittency filtering is increased. Reinforcing what was observed above, the transverse structure functions approach Kolmogorov scaling over all orders more rapidly than the longitudinal structure functions. 

Scaling of the structure functions is connected to the statistics of coarse-grained dissipation via the Obukhov-Kolmogorov refined similarity hypothesis~\cite{kolmogorov1962refinement}. Perhaps the most modern rationalization of the anomalous scaling of structure function and hence the coarse-grained dissipation statistics comes from the Parisi-Frisch multifrcatal formalism~\cite{Frisch-Parisi,Frisch_1995}. It is interesting to consider the effect of our intermittency filtering on the multifractal properties of filtered dissipation fields. 

Within the multifractal framework, the energy dissipation field $2\nu S_{ij}S_{ij}$ is conceived to be a multiplicative accumulation of a measure over the cascading process~\cite{Frisch_1995,meneveau1987multifractal,mandelbrot1989multifractal,o1993spatial}. This leads to an extremely patchy dissipation field with sharp spikes in isolated regions. It is then conjectured that the dissipation locally follows scalings of the form $r^{\alpha-1}$, $\alpha$ being the singularity strength related to the local H\"older exponent of the flow ($h=\alpha/3$), following the simple scaling argument, that if $\delta u \sim r^h$, then $\varepsilon \sim u^3/r \sim r^{3h-1}$. The scaling exponent in the range $[\alpha,\alpha + \mathrm{d}\alpha]$ is believed to hold over a fractal set of dimension $f(\alpha)$. Kolmogorov theory predicts a single space-filling scaling exponent (corresponding to $h=1/3$, which gives the $p/3$ scaling of structure function exponents $\zeta_p$). In the multifractal formalism, it is assumed that there are a range of singularity strengths $\alpha$ that the flow can assume, which are expressed over interwoven fractal sets of corresponding dimensions $f(\alpha)$. 

\begin{figure}
    \centering
    \includegraphics[width=1\linewidth]{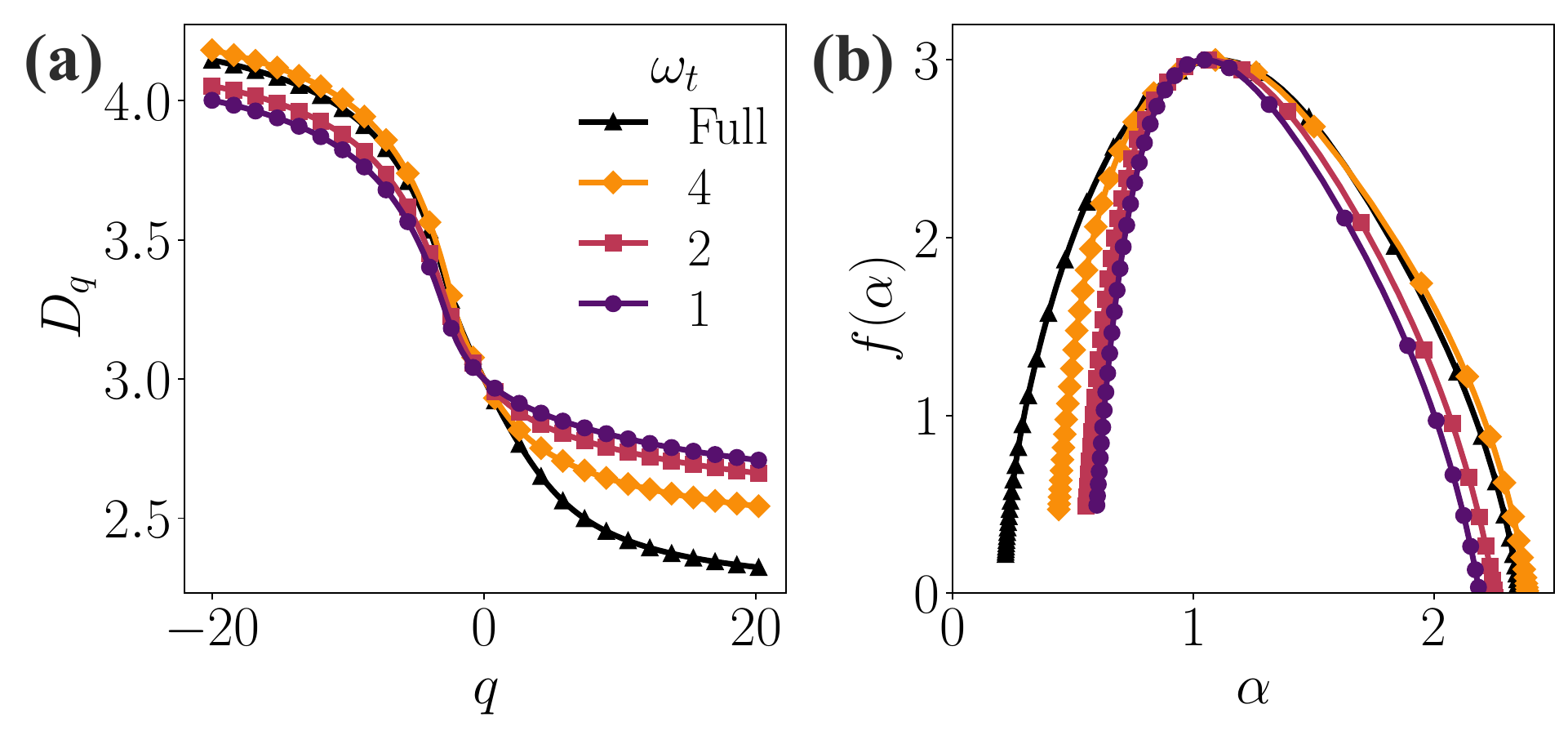}
    \caption{\textbf{Multifractal statistics.} \textbf{(a)} Generalized dimensions $D_q$ versus $q$ for the full turbulent dissipation field along with filtered dissipation fields over different levels of thresholding are shown. The spread of the distribution shrinks, with increased intermittency filtering. This is clearly reflected in the \textbf{(b)} singularity spectra $f(\alpha)$ versus $\alpha$. Increasing intermittency filtering clearly reduces the extent of \textit{singular} regions in the flow. This is seen from the reducing extent of the \textit{roughness} scaling exponents to the left-half of the $f(\alpha)$ curves, which corresponds to the $q>0$ half of $D_q$ (and is sensitive to extreme, large values of the field).}
    \label{fig4}
\end{figure}

\begin{figure*}
	\centering
	\includegraphics[width=1\linewidth]{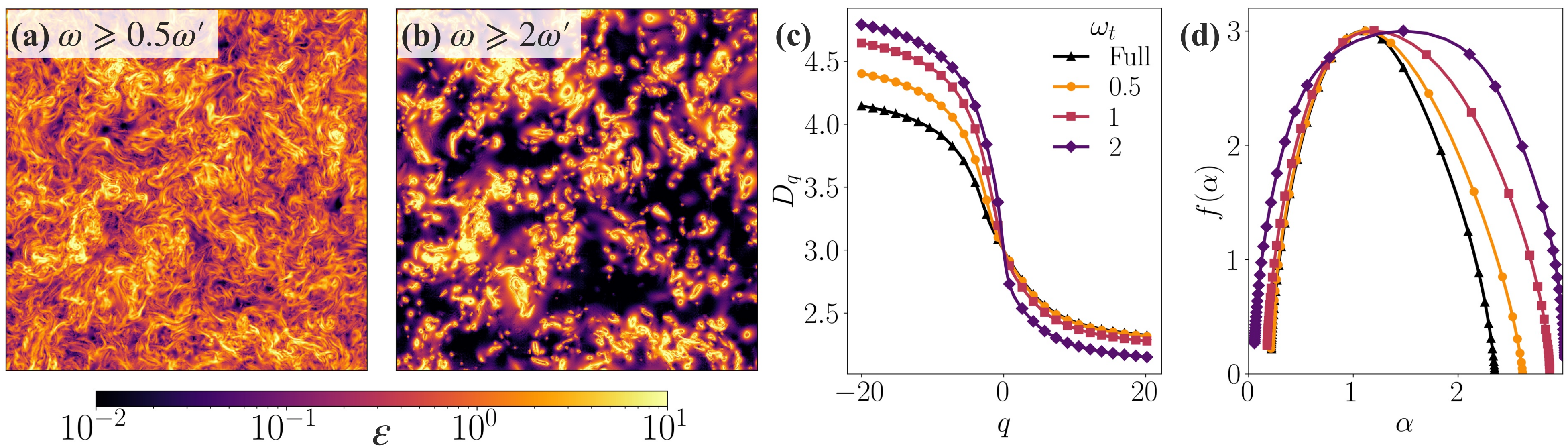}
	\caption{\textbf{Residual dissipation.} The residual dissipation field $\epsilon = 2\nu \widetilde{S}_{ij}^R\widetilde{S}^R_{ij}$ is shown for \textbf{(a)} $\omega \geqslant 0.5\omega^\prime$ and \textbf{(b)} $\omega \geqslant 2\omega^\prime$. The dissipation field in \textbf{(a)} looks rather similar to a regular turbulence dissipation field, while the field in \textbf{(b)} is very different, with large empty patches and small regions of strong dissipation, corresponding to regions of strong vorticity. Multifractal analysis gives the \textbf{(c)} generalized dimensions $D_q$ v/s $q$ and \textbf{(d)} singularity spectra $f(\alpha)$ v/s $\alpha$, for residual dissipation fields for various thresholds. This reveals that with increasing thresholding, the residual dissipation field becomes more multifractal, as the $f(\alpha)$ distributions spread in both the \textit{roughness} and \textit{smoothness} scaling exponent directions. The $\omega_t = 2\omega^\prime$ case shows a spectrum that begins to deviate from the base state of turbulence, and reveals a field organization that is more singular and lacks the background Kolmogorovean structure, which is centered at a strict peak around $h=\alpha/3=1/3$, as seen for the other cases.}
	\label{fig5}
\end{figure*}

Usually, the singularity spectrum $f(\alpha)-\alpha$ is obtained via a Legendre transform after the of generalized dimensions $D_q$ over $q$~\cite{grassberger1983characterization,Halsey_1986,hentschel1983infinite,procaccia1988universal,MS-JFM,MS-Nucl}. Here, the $D_q$ are obtained from the scaling of the partition function 
\begin{equation}
Z_q(\ell) = \sum_{N_\ell} \mathcal{E}(\ell)^q  \sim \ell^{D_q(q-1)} 
\end{equation}
over coarse-graining lengthscale $\ell$ (in the inertial range), and $\mathcal{E}(\ell)  = \displaystyle\sum_{r^\prime\in \ell}\varepsilon(r^\prime)$. From this, one gets: 
\begin{equation}
    D_q = \lim_{\ell\to 0} \frac{1}{q-1} \log{Z_q(\ell)} 
\end{equation}
From which it follows that
\begin{align}
    f(\alpha) = q\,\alpha - (q-1)(D_q-d+1) + d-1,\\
    \alpha(q) = \frac{\mathrm{d}}{\mathrm{d}q}\left[(q-1)(D_q-d+1)\right],
    \label{eq:alpha}
\end{align}
where $d=3$ is the embedding dimension of the measure field.

We start with a note about interpreting dissipation fields corresponding to the reconstructed velocity fields following our intermittency filtering procedure. The full dissipation field, can be written as:
\begin{align}
 2\nu S_{ij}S_{ij} &= 2\nu [\partial_i (\widetilde{u}_j + \widetilde{u}_j^R)/2 + \partial_j (\widetilde{u}_i + \widetilde{u}_i^R)/2]^2  \nonumber \\
 &= 2\nu [(\partial_i\widetilde{u}_j + \partial_j \widetilde{u}_i)/2 + (\partial_i\widetilde{u}^R_j+\partial_j\widetilde{u}^R_i)/2]^2 \nonumber \\
 &= 2\nu (\widetilde{S}_{ij} + \widetilde{S}_{ij}^R)^2
\end{align}

The pure background strain field, therefore can be seen to be $\widetilde{S}_{ij}$, which can be associated with a dissipation field $\varepsilon = 2\nu\widetilde{S}_{ij}\widetilde{S}_{ij}$. We study the behaviour of this field, while ignoring the residual dissipation field ($2\nu\widetilde{S}^R_{ij}\widetilde{S}^R_{ij}$) along with the more complex cross-term. 

In Fig.~\ref{fig3}(a)-~\ref{fig3}(c), we show the filtered dissipation fields for the velocity reconstructed after increasing levels of thresholding, at a representative crossection of the 3D dataset, with a common logarithmically spaced colorbar. As the level of filtering is increased by reducing the threshold, intense dissipation begins to be removed from the flow. This is also reflected in the probability distribution of the dissipation field, shown in Fig.~\ref{fig3}(d), where the dissipation $\varepsilon$ of the filtered velocity fields $\widetilde{\bf u}$ has been normalized by the standard deviation $\sigma_\varepsilon$. The tails of the pdf shrink with increased intermittency filtering. It is instructive to consider the field plots together with the pdf. Interestingly, a significant amount of dissipation extremes is already filtered out at the threshold of $\omega_t = 4\omega^\prime$. These are likely to be regions of high, locally-induced strain around cores of vortex-stretching. It is known that intense vorticity in turbulence is surrounded by self-generated sleeves of high strain, and hence high dissipation~\cite{picardo2019flow}. As the filtering is increased by reducing the threshold, the dissipation field becomes entirely quiescent in magnitude. Naturally, when $\omega_t = \omega^\prime$, vorticity is being filtered at a level to start altering the \textit{background} strain field in the flow. Remarkably, much like the kinetic energy field, the dissipation field reveals rich structure even at low magnitude levels. 

Generalized dimensions from the multifractal analysis of filtered dissipation are shown in Fig.~\ref{fig4}(a) and the corresponding singularity spectra in Fig.~\ref{fig4}(b). Note that, in general the peak of the singularity spectrum is around $\alpha=1$, which coincides with the Kolmogorov prediction $h = \alpha/3=1/3$. To the left of $\alpha=1$ denotes the singular extent of the spectrum ($q>0$), corresponding to the \textit{rough} regions of the flow field (intense dissipation), which we refer to as the \textit{roughness} scaling exponents. The right half of the spectrum ($q<0$) corresponds to the smoother regions of the flow (quiescent regions), referred to as the \textit{smoothness} scaling exponents.

To understand the effect of intermittency filtering, we focus on the left-half of the $f(\alpha)$ curves in Fig.~\ref{fig4}(b). We first note that the largest extent of $f(\alpha)$ in the left-half, naturally, is reached for the full flow. Interestingly, there is a clear trend upon increasing the intermittency filtering. The distribution of $f(\alpha)$ shrinks towards the peak as vorticity is increasingly filtered out. The shrink is seen to be more than half of the full field multifractal spectrum. We note that even with the highest level of filtering considered, the distribution does not become monofractal, as predicted by Kolmogorov theory. The variations within the field, even where the flow is relatively smooth, yields a broad distribution of $f(\alpha)$ to the right of the peak. 

It is natural to now ask what is the structure of the residual fields. We recap that these are velocity fields $\widetilde{\bf u}^R$ obtained by inverting the vorticity field conditioned over regions where $\omega \geqslant \omega_t$. Typically, when $\omega_t$ is high, there are so few and far apart regions of the vorticity field which contribute to generating the velocity, that there is no large scale organization in the reconstructed flow field. As such, it is difficult to find a scaling range in the structure functions, which is why we do not show them here. 

It is useful, instead, to probe the structure of the residual dissipation fields via multifractal analysis, which leads to interesting insights. In Fig.~\ref{fig5}(a) we show the residual dissipation field $\varepsilon = 2\nu \widetilde{S}_{ij}^R\widetilde{S}^R_{ij}$ corresponding to the residual velocity field $\widetilde{\bf u}^R$, for $\omega_t = 0.5\omega^\prime$ (so the velocity field is constructed from the range $\omega \geqslant 0.5\omega^\prime$). This field looks much like the (full) dissipation field typically observed in turbulence, which is understandable due to the low threshold. Interestingly, in Fig.~\ref{fig5}(b) for $\omega_t = 2\omega^\prime$, the field is remarkably different. There are large voids of almost neglibile dissipation, interspersed by small islands of intense dissipation, which naturally correspond to neighbourhoods of high vorticity (in the range $\omega \geqslant 2\omega^\prime$). In a way, this does not much resemble a turbulence dissipation field in terms of structures and organization, however, the variation in residual dissipation appears more extreme.

This is very clearly reflected in the multifractal analysis. Fig.~\ref{fig5}(c) shows the generalized dimensions $D_q$ v/s $q$ for the residual dissipation fields over different levels of thresholding. The spread of $D_q$ increases with the threshold $\omega_t$, reflecting a broader range of singularity strengths. This is confirmed from Fig.~\ref{fig5}(d) showing the $f(\alpha)-\alpha$ singularity spectra. The residual dissipation field becomes more singular, as seen from the increase in both the \textit{spread} of $f(\alpha)$ and the \textit{extent} of $\alpha$, on either side of the peak of the distribution. Higher thresholds indeed make the residual field attain more extreme $\alpha$ values. What is also evident is that for $\omega_t = 2\omega^\prime$, the peak and shape of the distribution changes from the others which remain pinned around $\alpha = 1$. In turbulence, this reflects that the base field is Kolmogorovean, with the scaling exponent $h = \alpha/3 = 1/3$, around which there is a finite distribution of $\alpha$, due to intermittency. The residual field for $\omega_t = 2\omega^\prime$ begins to deviate from this underlying Kolmogorovean structure entirely, reflecting that the flow generated by $\omega \geqslant 2\omega^\prime$ does not conform to turbulence scalings. This is essentially a large, quiescent field with small, localized intense vortices, which is distinct from the multiscale structure of turbulent flow. At $\omega \geqslant 4\omega^\prime$ (not shown), it becomes difficult to find a scaling range in the partition function moments, and the computed singularity spectrum becomes even more deformed and broad. This again is a reminder that it is indeed mild vorticity that generates the overall Kolmogorov turbulence. We expect \textit{local multifractality}~\cite{mukherjee2024turbulent} to also be significantly altered by the filtering, and for filtered and residual dissipation to show distinct behaviour, which we leave for future work to explore.

\begin{figure}
    \centering
    \includegraphics[width=\linewidth]{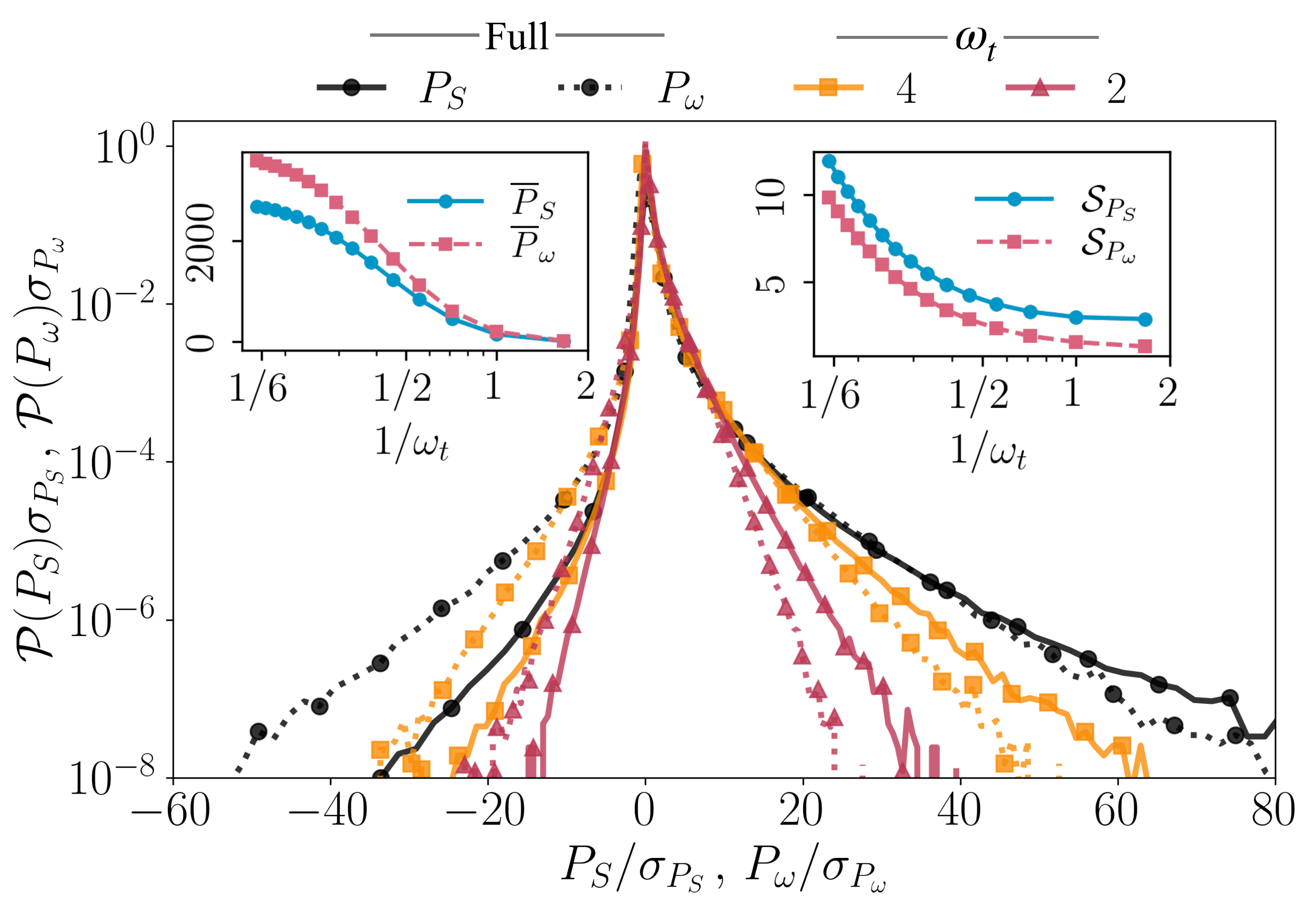}
    \caption{\textbf{Strain self-amplification and vortex stretching.} Probability distributions of strain-self amplification $P_S = -S_{ij}S_{jk}S_{ki}$ (solid lines) and the vortex-stretching term $P_\omega = \omega_i \omega_j S_{ij}$ (dotted lines) for different levels of intermittency filtering. Black lines refer to the full field and the colors refer to different thresholds (see legends above). Insets show the variation of the mean (top left) and skeweness (top right) of the $P_S$ and $P_\omega$ distributions.}
    \label{fig6}
\end{figure}

To make a final link between our filtering procedure, it's effect on flow structures and the changes to the dissipation field, we consider the transport equation for strain energy $S_{ij}S_{ij}$ which is given as

\begin{equation}
    \frac{1}{4}\frac{{\rm D} S_{ij}S_{ij}}{{\rm D}t} = -S_{ij}S_{jk}S_{ki} - \frac{1}{4}\omega_i \omega_j S_{ij} - S_{ij}\Pi_{ij} + \nu S_{ij}\nabla^2 S_{ij}
\end{equation}

where $\Pi_{ij}$ is the pressure Hessian tensor. The first two terms on the right of the equation are namely the strain-self amplification $P_S = -S_{ij}S_{jk}S_{ki}$ and the vortex-stretching term $P_\omega = \omega_i \omega_j S_{ij}$ which leads to enstrophy production in the vorticity transport equation. Both these terms contribute to intensifying the dissipation field~\cite{Johnson_2021,buaria2022generation}, while the dynamical processes, and hence flow structures, behind them are distinct. We recall that vortex-stretching occurs most strongly from the alignment of the local vorticity with the largest stretching direction of the \textit{background} strain field induced non-locally in a Biot-Savart sense, which does not include the velocity field induced by the local vorticity~\cite{hamlington2008local}. So, while strain and vorticity amplification are both susceptible to the background velocity field, our filtering approach will directly influence vortex stretching more prominently. 

Fig.~\ref{fig6} shows the probability distribution of $P_S$ (solid lines) and $P_\omega$ (dotted lines), for different levels of thresholding. We find that the distribution of $P_\omega$ has more extreme tails towards the negative side of the distribution. Even a small amount of filtering with $\omega_t = 4\omega^\prime$ significantly shortens the $P_\omega$ tail, while the $P_S$ tail moves only by a small factor. With slightly more filtering, the $P_\omega$ distribution becomes closer to $P_S$. We further quantify the effect of our filtering by looking at the change of the mean (top left inset) and skewness (top right inset) of $P_S$ and $P_\omega$. We note that there is a relation between the mean of there two quantities in turbulence, $\overline{P}_s = (3/4) \overline{P}_\omega$~\cite{betchov1956inequality}. We find in Fig.~\ref{fig6} (top left inset) that with increasing thresholding, both $\overline{P}_\omega$ and $\overline{P}_S$ fall to small values, and the ratio $\overline{P}_s/\overline{P}_\omega$ goes from $0.75$ ($\omega_t = 6\omega^\prime$) to $0.73$ (at a thresholding of $\omega_t = 0.5\omega^\prime$). The skewness Fig.~\ref{fig6} (top right inset) of both the quantities also decreases, while their ratio $\mathcal{S}_{P_s}/\mathcal{S}_{P_\omega}$ goes from $\approx 1$ to $\approx 2$, showing that $\mathcal{S}_{P_\omega}$ falls faster than $\mathcal{S}_{P_s}$. This shows that flow structures influence different aspects of the budget equation differently. 

\begin{figure}
    \centering
    \includegraphics[width=1\linewidth]{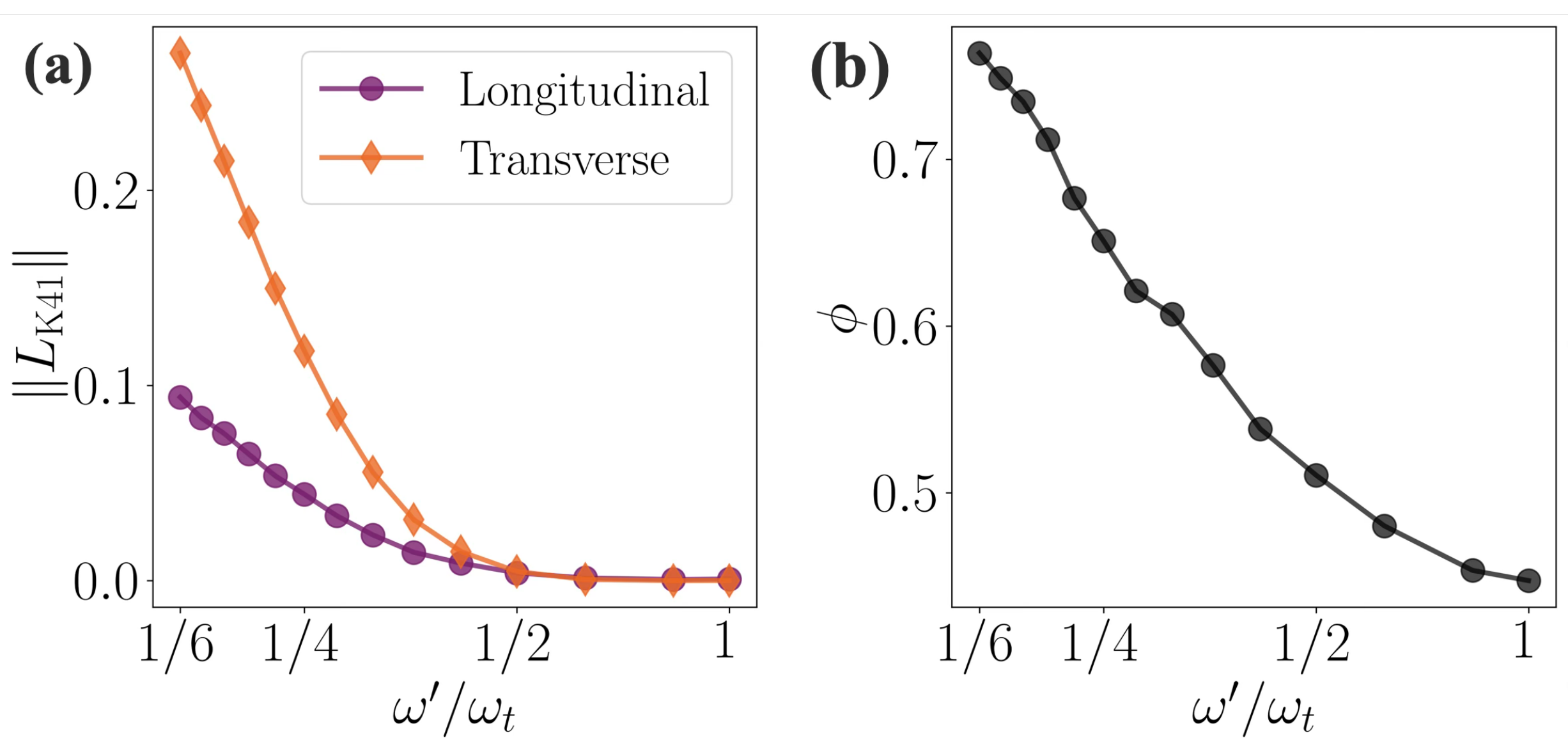}
    \caption{\textbf{Approach to Kolmogorov Scaling.} \textbf{(a)} Deviation from Kolmogorov scaling is quantified as norm of scaling exponents $\zeta_p$ deviating from $p/3$, given as $||L_{\rm K41}|| = \sum_p(\zeta_p - p/3)^2$, over the inverse of the filtering threshold $1/\omega_t$. As the filtering is increased (left to right on the x-axis), the fields approach Kolmogorov scaling. The steps of the approach are steeper for the transverse structure function exponents, $\zeta^T_p$, in comparison to the longitudinal exponent $\zeta^L_p$. \textbf{(b)} The shrink in the singular part of the multifractal spectrum is shown as $\phi = \alpha_{\rm peak} - \alpha_{\rm min}$ (range of \textit{roughness} scaling exponents), which reflects a linear shrinkage in the singularity strengths with increasing intermittency filtering.}
    \label{fig7}
\end{figure}

Finally, we close with a summary of our main findings in Fig.~\ref{fig7}, which shows the approach to Kolmogorov scaling with increased intermittency filtering. Fig.~\ref{fig7}(a) shows the deviation of the structure function scaling from the Kolmogorov prediction as the norm $\lVert L_{\rm K41}\rVert = \sum_p ( \zeta_p - p/3 )^2$, for both the longitudinal and transverse structure functions over a wide range of intermittency filtering thresholds, shown as $\omega^\prime/\omega_t$ (such that increasing filtering is from left to right). Clearly, beyond $\omega_t = 2\omega^\prime$, both the structure functions become essentially Kolmogorovean, showing that the velocity field generated by the vorticity upto $2\omega^\prime$ obeys Kolmogorov scaling. The approach to Kolmogorov scaling happens with larger steps for transverse structure functions, as opposed to longitudinal. This reveals an essential link of anomalous scaling with flow structures---the filtering of intermittent vorticity, which is associated with local swirling flow, more prominently influence transverse scaling. Fig.~\ref{fig7}(b) shows the extent of the singular part of the multifractal spectrum $f(\alpha)$ versus $\alpha$, calculated as the range of \textit{roughness} scaling exponents $\phi = \alpha_{\rm peak} - \alpha_{\rm min}$, which is the range of $\alpha$ values to the left of the peak of the curve. For the unfiltered flow, the distribution is broad in this region. But as intermittency filtering increases, we find that there is an asymptotic reduction in $\phi$. 

\section{Conclusions}\label{conclusions}
Explaining intermittency in turbulence, in the form of both extreme vorticity and dissipation, has remained beyond the ambit of a first-principles theoretical framework. Intermittency is now known synonymously with its effects on scaling behaviour, namely anomalous scaling of structure functions, corrections to energy spectrum scaling and multifractality. While the qualitative effects of intermittent flow structures on these measurements is not contested, there is no singular approach to address this quantitatively. In this work, we introduce a simple intermittency filtering procedure to selectively remove the influence of a chosen range of the vorticity field in kinematically inducing the velocity field. This is done by limiting the vorticity field to below a chosen threshold $\omega_t$, and inverting the Biot-Savart law spectrally to obtain filtered velocity fields $\widetilde{\bf u}$. This procedure can be seen as an exact disentangling of the different vorticity contributions that build up the full velocity field. Such an approach has been used before to study the Biot--Savart composition of coherent structures~\cite{mukherjee2022correlation} and to unravel vorticity-strain dynamics~\cite{hamlington2008local}.

We study the statistics of filtered velocity fields with successively higher thresholds. We find that the filtered velocity field retains rich structure even when it is generated only by $\omega < \omega^\prime$. The scaling of energy spectra persists, and in fact is seen to \textit{improve} with a flattening of the bottleneck. This is a first hint that intermittent structures also concentrate sufficient energy at small scales to produce a bump in the energy spectra. Structure function scaling is found to approach Kolmogorov scaling of $\zeta_p = p/3$, both for transverse and longitudinal structure functions, with increasing thresholding. The role of intermittent flow structures again plays out in the more rapid approach of transverse scaling exponents to their classical $p/3$ values, than longitudinal exponents. We conjecture that it must be the local swirling flow regions generated by intense vorticity that selectively influence transverse structure functions more, due to the propensity of anti-aligned velocity vectors across vortex cores. 

Multifractal analysis of filtered dissipation $\varepsilon = 2\nu \widetilde{S}_{ij}\widetilde{S}_{ij}$ reveals that intermittency filtering, naturally, shrinks the range of roughness scaling exponents in the flow (which are the exponents to the left of the peak of the $f(\alpha)$ distribution). As intermittency is removed, the flow becomes smoother. There is in fact an amplification of the smoothness exponents (those to the right of the $f(\alpha)$ peak). We also probe the statistics of the residual dissipation field $\varepsilon = 2\nu \widetilde{S}_{ij}^R\widetilde{S}_{ij}^R$. We find that the structure of these fields is significantly different from turbulence, specially when the residual fields are constructed from vorticity above medium to high thresholds ($\omega_t \geqslant 2\omega^\prime$). The singularity spectrum breaks away from its peak at $\alpha=1$ and becomes broader, revealing a more multifractral residual field. Together, these results show that the Kolmogorov theory prediction of a single scaling exponent, $h=\alpha/3=1/3$, is the central scaffold around which a range of singularity exponents are distributed, as a consequence of intermittency. Velocity fields generated by intermittent vorticity alone are not even turbulent, in this sense. We also note that with our Biot--Savart procedure is an operational filtering, while the dynamics of turbulence is not influenced, as in the case of other intermittency removal procedures like the decimated Navier--Stokes approach. Consequently, the filtered dissipation fields here show a richer structure than under decimation. We quantify the approach to Kolmogorov scaling over our filtering threshold, as a normed distance $\lVert L_{\rm K41} \rVert$ from Kolmogorov scaling, and extent of roughness exponents $\phi$. 

Our work shows a simple method to surgically remove intermittency from turbulence at any chosen level. This procedure also allows a direct interpretation of the filtering in terms of removing intense vortices, that are swirling velocity structures surrounding strong vorticity cores. The method allows retaining the background turbulence velocity field. This opens up new ways of addressing long standing questions about dynamics, structures, scalings and the estimation of universal measures like the Kolmogorov constant.

\section{Data Availability Statement}
All analysis codes have been shared on a public GitHub repository~\cite{mukherjee_intermittency}. Turbulence field data can be shared upon reasonable request from the authors.

\section{Acknowledgments}
We thank Samriddhi Sankar Ray for continued discussions. RM acknowledges the support of
the DAE, Government of India, under Projects No. RTI4019 and No. RTI4013. The simulations and analysis were performed on ICTS-TIFR HPC cluster \textit{Contra}. SM would like to thank ICTS-TIFR, Bengaluru for support and hospitality. SM acknowledges the Govt. of India grant ANRF/ECRG/2024/002467/ENS and the IITK Initiation Grant IITK/ME/2024316.

\bibliographystyle{apsrev4-2}
\bibliography{reference}

\end{document}